\documentclass{PoS}
\usepackage{epsfig}
\usepackage{latexsym}
\usepackage{amssymb}
\usepackage{amsmath}
\usepackage{graphicx}
\usepackage{dcolumn}
\usepackage{bm}

\renewcommand{\vec}[1]{\boldsymbol{#1}}
\newcommand{\be}{\begin{equation}}
\newcommand{\ee}{\end{equation}}
\newcommand{\ba}{\begin{eqnarray}}
\newcommand{\ea}{\end{eqnarray}}
\newcommand{\bi}{\begin{itemize}}
\newcommand{\ei}{\end{itemize}}
\newcommand{\tr}{{\rm Tr\,}}

\newcommand{\eq}{Eq.~}

\newcommand{\fig}{Fig.~}

\newcommand{\la}{\label}

\title{Thermal field theories and shifted boundary conditions}

\ShortTitle{$\dots$ shifted boundary conditions}

\author{\speaker{Leonardo Giusti}\\
Dipartimento di Fisica, Universit\`a di Milano--Bicocca,\\
and INFN, sezione di Milano--Bicocca,\\
I-20126 Milano, Italy\\
E-mail: \email{Leonardo.Giusti@mib.infn.it}}

\author{Harvey B. Meyer \\
PRISMA Cluster of Excellence, \\ 
Institut f\"ur Kernphysik and Helmholtz Institute Mainz\\
Johannes Gutenberg-Universit\"at Mainz\\
D-55099 Mainz, Germany\\
E-mail: \email{meyerh@kph.uni-mainz.de}}

\abstract{
The analytic continuation to an imaginary velocity of the canonical 
partition function of a thermal system expressed in a moving frame 
has a natural implementation in the Euclidean path-integral formulation 
in terms of shifted boundary conditions. The Poincar\'e invariance  
underlying a relativistic theory implies a dependence of the 
free-energy on the compact length $L_0$ and the shift ${\vec \xi}$ only through 
the combination $\beta=L_0(1+{\vec \xi}^2)^{1/2}$. This in turn implies 
that the energy and the momentum distributions of the thermal theory are
related, a fact 
which is encoded in a set of Ward identities among the correlators of 
the energy-momentum tensor. The latter have interesting applications in
lattice field theory: they offer novel ways to compute
thermodynamic potentials, and a set of identities to renormalize 
non-perturbatively 
the energy-momentum tensor. At fixed bare parameters the shifted boundary 
conditions also provide a simple method to vary the temperature in much
smaller steps than with the standard procedure.}

\FullConference{31st International Symposium on Lattice Field Theory - LATTICE 2013\\
		July 29 - August 3, 2013\\
		Mainz, Germany}

\begin{document}

\section{Introduction}
\vspace{-0.25cm}

A relativistic thermal field theory can be formulated 
in the Euclidean path integral formalism by imposing 
on the fields periodic boundary conditions\footnote{To avoid unessential 
technical complications we restrict ourselves to bosonic theories in 
this presentation.} in the compact direction up to a shift 
${\vec \xi}$ in the spatial 
directions~\cite{Giusti:2010bb,Giusti:2011kt,Giusti:2012yj}
\be\label{eq:sbc}
\phi(L_0, \vec x) = \phi(0, \vec x - L_0\, \vec \xi)\; .  
\ee
The free-energy density can be defined as usual
\be\label{eq:freeE} 
f(L_0,\vec \xi) = - \frac{1}{L_0 V} 
\ln Z(L_0,\vec \xi)\; ,
\ee
where $Z(L_0,\vec \xi)$ is the partition function, 
and $V$  is the spatial volume. In the thermodynamic 
limit the invariance of the dynamics under the SO(4)
group implies
\be\la{eq:fEucl.eq.fcan}
f(L_0,\vec \xi) = f(L_0\sqrt{1+\vec \xi^2},\vec 0)\; ,
\ee
i.e. the free energy is independent on the angles between 
the time and the space directions, while it depends on 
the length of the compact direction $\beta= L_0 \sqrt{1+\vec \xi^2}$ 
which fixes the inverse temperature of the system. 
This redundancy implies that the total energy and momentum 
distributions of the thermal theory are related, and interesting 
Ward identities (WIs) follow. As a result thermodynamic potentials, 
which are usually extracted from the free energy itself and from 
the energy distribution of the theory, can be extracted from the 
momentum distribution as well. 

These ideas find interesting applications when a theory is discretized
on the lattice, where the momentum distribution is easier to access 
in presence of a non-zero shift in the boundary conditions~\cite{Giusti:2010bb}.
In this talk we review the derivation of Eq.~(\ref{eq:fEucl.eq.fcan}), 
of the WIs that it implies, and we show some potentially interesting 
applications on the lattice. A full-fledged discussion on this topic 
as well as the unexplained notation can be found in the original 
references~\cite{Giusti:2010bb,Giusti:2011kt,Giusti:2012yj}.
\vspace{-0.25cm}

\section{Euclidean theory with shifted boundary conditions\la{sec:tftsbc}}
\vspace{-0.25cm}

Consider a quantum field theory defined on $\mathbb{R}^4$, an orthonormal 
basis, and $4$ linearly independent primitive vectors 
$ v^{(\mu)}$ ($\mu=0,1,\dots,3$). The latter can be represented by a 
primitive matrix $V\in {\rm GL}(4,\mathbb{R})$ whose columns are 
the components of $v^{(\mu)}$ in the orthonormal basis. For a given point 
labeled with the coordinates $x_\mu$, the field is identified at all points 
with coordinates
\be
x_\mu + V_{\mu{\nu}} m_{\nu},\qquad m_\nu\in\mathbb{Z}\; ,
\ee
i.e. we impose generalized periodic boundary conditions (GPBCs). 
The shifted boundary conditions which implement the partition 
function in Eq.~(\ref{eq:freeE}) are a special case of GPBCs. By defining 
the primitive cell as usual
\be
\Omega = \Big\{ x \in \mathbb{R}^4\,|\;  
       x_\mu = V_{\mu{\nu}} t_{\nu},\;
            0\leq t_\mu  < 1 \Big\}\; ,
\ee
6 parameters specify the orientation of the cell 
while 10 fix its geometry. For a Lorentz-invariant theory in a finite volume, 
the most general relation between two primitive matrices
$V$ and $W$ corresponding to a theory with 
two different sets of GPBCs and equal partition functions, is given by 
\be\la{eq:equiv}
 W = \Lambda\, V M, \qquad \Lambda\in{\rm SO}(4),
\quad M\in {\rm SL}(4,\mathbb{Z})\; .
\ee
The matrix $M$ modifies the geometry of the primitive cell, while
$\Lambda$ modifies its orientation. The freedom to choose the former
is a property of periodic boundary conditions, the freedom to choose
the latter is a property of the SO($4$) invariance of the
theory which in turn allows one to derive \eq(\ref{eq:fEucl.eq.fcan}) and 
the corresponding WIs. The partition function
\be\la{eq:defZEucl}
Z(V_{\rm sbc}) = 
\tr\{e^{-L_0(\widehat H-i\vec\xi\cdot\widehat{\vec P})}\}\; , \qquad
V_{\rm sbc} = \left(\begin{array}{c@{~~}c@{~~}c@{~~}c}
L_0 &  0  & 0 & 0  \\
L_0\xi_1  & L_1 & 0 & 0 \\
L_0\xi_2  & 0 & L_2 & 0 \\
L_0 \xi_3 & 0 & 0 & L_3 
\end{array}   \right)\; , 
\ee
can be expressed as a Euclidean path integral with the fields satisfying 
standard periodic boundary conditions in the spatial directions, and 
the shifted boundary conditions in Eq.~(\ref{eq:sbc}). By defining 
\be
V_1 = M^{-1} R\, V_{\rm sbc} M = \left(\begin{array}{c@{~~~}c@{~~~}c@{~~~}c}
L_1\gamma_1 & 0 & 0 & 0 \\
-L_1\gamma_1\xi_1 & L_0/\gamma_1 & 0 & 0 \\
0 & L_0\xi_2 & L_2 & 0 \\
0 & L_0\xi_3 & 0 & L_3 \\
\end{array} \right)
\ee
with
\be
R=\left(\begin{array}{c@{~~~}c@{~~~}c@{~~~}c}
\gamma_1 & \gamma_1 \xi_1 & 0 & 0 \\
-\gamma_1 \xi_1 & \gamma_1 & 0 & 0 \\
0 & 0 & 1 & 0 \\
0 & 0 & 0 & 1 \\
\end{array}   \right),
\qquad
M=\left(\begin{array}{c@{~~~}c@{~~~}c@{~~~}c}
0 & 1 & 0 & 0 \\
-1 & 0  & 0 & 0 \\
0 & 0 & 1 & 0 \\
0 & 0 & 0 & 1 \\
\end{array}   \right),
\ee
and $\gamma_k = \big(1+\xi_k^2\big)^{-{1}/{2}}$, we conclude 
that $Z(V_{\rm sbc}) = Z(V_1)$. We first focus on the 
case $\xi_2=\xi_3=0$, and later use the 
SO(3) rotation symmetry to generalize the result to a generic shift
vector. The partition function can be interpreted in terms of the 
states that propagate in the direction given 
by the first column of $V_1$.  In the thermal field theory language, the 
latter are the eigenstates of the `screening' Hamiltonian 
$\widetilde H$, which acts on states living on a slice of dimensions 
$(L_0/\gamma_1)\times L_2 \times L_3$
with ordinary periodic boundary conditions. Their spectrum yields the
spatial correlation lengths of the thermal system at inverse
temperature $(L_0/\gamma_1)$. The partition function can thus 
be written as 
\be\la{eq:Zscreen}
Z(V_1) = 
\tr\Big\{\exp{-L_1\gamma_1(\widetilde H + i {\xi_1}\widetilde\omega)}\Big\},
\ee
where $\widetilde\omega$ is the momentum operator along the primitive vector of
length $(L_0/\gamma_1)$.  Its eigenvalues are the Matsubara frequencies
$\omega_n = \gamma_1 \frac{2 \pi n}{L_0}$, $n\in\mathbb{Z}$.  Assuming that
the Hamiltonian $\widetilde H$ has a translationally invariant vacuum
and a mass gap, the right-hand side of \eq(\ref{eq:Zscreen})
becomes insensitive to the phase in the limit $L_1\to\infty$ at fixed
$\xi_1$ (with exponentially small corrections, see Ref.~\cite{Giusti:2012yj}). 
This in turn 
implies that the free energy densities associated with $V_{\rm sbc}$ and 
$\mbox{diag}(L_1\gamma_1, L_0/\gamma_1,L_2 ,L_3)$ are equal. 
Thanks to the invariance of the infinite-volume theory under 
three-dimensional rotations, this result extends to a generic imaginary 
velocity $\vec\xi$. In the thermodynamic limit the net effect
of the generic shift $\vec\xi$ is thus to lower the temperature from $1/L_0$ to 
$1/\beta=1/(L_0\sqrt{1+{\vec\xi}^2})$, i.e.\ we have 
proved \eq(\ref{eq:fEucl.eq.fcan}). The latter
is consistent with modern thermodynamic arguments on the Lorentz 
transformation of the temperature and the free energy ~\cite{Ott,Arz} 
(the issue has been debated for a long time, see Ref.~\cite{Prz} for a 
recent discussion), see Ref.~\cite{Giusti:2012yj} for more details.
\vspace{-0.25cm}

\section{Ward identities for the total energy and momentum}
\vspace{-0.25cm}

The relation (\ref{eq:fEucl.eq.fcan}) is the source of certain WIs for 
the energy-momentum tensor which can be generated in a quasi-automated 
fashion by deriving the free-energy density with respect to $L_0$ and 
$\xi_k$. By remembering that the cumulants of the total momentum 
distribution can be written as
\be\label{eq:cum1}
k_{\{2 n_1, 2 n_2, 2 n_3\}} \!\equiv\! \frac{1}{V}\langle {\widehat P}_1^{2 n_1} 
{\widehat P}_2^{2 n_2} {\widehat P}_3^{2 n_3} \rangle_c \! 
=\! \frac{(-1)^{n_1+n_2+n_3+1}}{L_0^{2 n_1+2 n_2+2 n_3-1}}
\frac{\partial^{2n_1}}
{\partial \xi_1^{2n_1}} \frac{\partial^{2n_2}}{\partial \xi_2^{2n_2}}
\frac{\partial^{2n_3}}{\partial \xi_3^{2n_3}}
f(L_0,\vec \xi)\Big|_{\vec \xi=0},\!\!\!\!
\ee
in the thermodynamic limit a plethora of Ward identities among on-shell 
correlators of the total momentum and/or energy are derived by inserting 
\eq(\ref{eq:fEucl.eq.fcan}) in (\ref{eq:cum1}). By choosing 
$\vec \xi=\{\xi_1,0,0\}$, it is straightforward to derive the master 
equation
\be\label{eq:fnd1}
\frac{k_{\{2 n, 0, 0\}}}{L_0} =(-1)^{n+1}\, (2n-1)!!\,
\Big\{\frac{1}{L_0}\frac{\partial}{\partial L_0} \Big\}^n f(L_0,\vec \xi)\Big|_{\vec \xi=0}  
\qquad n=1,2,\dots \; . 
\ee
If we define $c_1 \equiv e-f$ and recall that the higher cumulants of the 
total energy distribution are given by
\be\la{eq:cn}
c_n \equiv \frac{1}{V}\, \langle\, \widehat H^n\, \rangle_c = (-1)^{n+1}\left[ n\, \frac{\partial^{n-1}}{\partial L_0^{n-1}}
+ L_0\, \frac{\partial^n}{\partial L_0^n} \right]
f(L_0,\vec \xi)\Big|_{\vec \xi=0} \quad n=2,3\dots\; ,
\ee
it is clear that there is a linear relation among 
$c_1,\dots, c_n$ and the $n$ first derivatives of 
the free-energy density. Since \eq(\ref{eq:fnd1}) gives the 
$k_{\{2 n, 0, 0\}}$ as linear combinations of the very same 
derivatives, the relation reads
\be\la{eq:k2ncn}
k_{\{2n,0,0\}} = \frac{(2n-1)!!}{(2 L_0^2)^n}\,\sum_{\ell=1}^n 
\frac{(2n-\ell)!}{\ell!(n-\ell)!}\, {(2 L_0)^\ell\, c_\ell }\, ,
\ee
Up to $n=4$ we obtain
\ba\label{eq:nice}
L_0\, k_{\{2,0,0\}\,} &=& c_1\;,\nonumber\\
L_0^3\, k_{\{4,0,0\}}\, &=& 9\, c_1 + 3\,L_0\,  c_2\;,\\
L_0^5\, k_{\{6,0,0\}} &=& 225\, c_1 + 90\,L_0\,  c_2 + 
15\, L_0^2\, c_3\;,\nonumber\\
L_0^7\, k_{\{8,0,0\}} &=& 11025\, c_1 + 4725\, L_0\,  c_2 + 
1050\,L_0^2\,  c_3 + 105\,L_0^3\,  c_4\;.\nonumber
\ea
If we remember that in the Euclidean
$\langle T_{00} \rangle = -e$, $\langle T_{kk} \rangle = p$, and 
$\widehat P_k \rightarrow -i {\overline T}_{0k}$, where 
${\overline T}_{\mu\nu}(x_0) = \int d^3 x\, T_{\mu\nu}(x)$ with 
$T_{\mu\nu}$ being the energy-momentum field of the theory,
Eqs.~(\ref{eq:nice}) can also be written as 
\ba\label{eq:contWIs}
L_0\, \langle {\overline T}_{01}\, T_{01}\rangle_c 
& = & \langle T_{00} \rangle - \langle T_{11} \rangle \; ,\nonumber\\[0.125cm]
L_0^3\, \langle {\overline T}_{01}\, {\overline T}_{01}\, 
                {\overline T}_{01}\, T_{01}\rangle_c & = &
9\, \langle T_{11} \rangle - 9\, \langle T_{00} \rangle 
+ 3\, L_0\,  \langle {\overline T}_{00} T_{00} \rangle_c\; ,\\
& \dots & \nonumber
\ea 
where in each correlator the energy-momentum fields are inserted  
at different times. These relations show that in a relativistic 
thermal theory the total 
energy and momentum distributions are related. The thermodynamics 
can thus be studied either from the energy or from the momentum 
distribution.
\vspace{-0.25cm}

\subsection{Ward identities in presence of a non-zero shift}
\vspace{-0.25cm}
If now the system is boosted by choosing 
$\vec \xi \neq 0$, standard parity is softly broken by the boundary 
conditions in the compact direction, odd derivatives in the $\xi_k$ 
do not vanish anymore, and new interesting WIs hold. By deriving 
once with respect to $L_0$ and $\xi_k$, it is easy to 
obtain the first non-trivial relation
\be\label{eq:WIodd}
\langle T_{0k} \rangle_{\vec\xi} = \frac{\xi_k}{1-\xi_k^2} 
\left\{\langle T_{00} \rangle_{\vec\xi}  
- \langle T_{kk} \rangle_{\vec \xi}\,\right\}\; .
\ee
An interesting consequence of this equation is that the 
entropy density $s$ of the system at the inverse temperature 
$\beta=L_0\sqrt{1+{\vec \xi}^2}$ is given by
\be\label{eq:spv}
s= - \frac{L_0}{\gamma^3\xi_k}\, \langle T_{0k} \rangle_{\vec\xi}\; ,
\ee 
where $\gamma=1/\sqrt{1+\vec\xi^2}$. Remarkably the entropy density
can be obtained directly from the vacuum expectation value of the 
off-diagonal component $T_{0k}$ of the energy-momentum tensor.
Ward identities among correlators with more fields can easily be 
obtained by considering higher order derivatives in $L_0$ and $\xi_k$.
For instance by deriving two times with respect to the shift components
we obtain
\be\label{eq:wipv}
\langle T_{0k}\rangle_{\vec\xi}  = 
\frac{L_0 \xi_k}{2}\,  \sum_{ij}\,
\left\langle {\overline T}_{0i}\,  T_{0j} \right\rangle_{\vec\xi,\, c}\, 
\left[ \delta_{ij} - \frac{\xi_i\, \xi_j}{\vec\xi^2}\right]\; . 
\ee
By combining Eqs.~(\ref{eq:spv}) and (\ref{eq:wipv}), the entropy 
density can also be computed as 
\be\label{eq:stwopt}
\displaystyle
s^{-1} =  -  
\frac{\gamma^3}{2}\,  \sum_{ij}\,
\frac{\left\langle{\overline T}_{0i}\,  T_{0j} \right\rangle_{\vec\xi,\, c}}
{\langle T_{0i} \rangle_{\vec\xi} \langle T_{0j}\rangle_{\vec\xi}}\, \xi_i \xi_j 
\,\Big[ \delta_{ij} - \frac{\xi_i \xi_j}{\vec \xi^2}\Big]\; 
,
\ee
and the analogous expression for the specific heat reads 
\be\label{eq:cvtwopt}
\displaystyle
\frac{c_v}{s^2} =  -  
\frac{\gamma^3}{2}\,  \sum_{ij}\,
\frac{\left\langle{\overline T}_{0i}\,  T_{0j} \right\rangle_{\vec\xi,\, c}}
{\langle T_{0i} \rangle_{\vec\xi} 
\langle T_{0j}\rangle_{\vec\xi}}\, \frac{\xi_i \xi_j}{\vec\xi^2} 
\,\Big[ (1-2\vec\xi^2)\delta_{ij} - 3 \frac{\xi_i \xi_j}{\vec \xi^2}\Big]\; .
\ee
\vspace{-0.75cm}

\section{Applications on the lattice\la{sec:appl}}
\vspace{-0.25cm}
The shifted boundary conditions discussed so far
provide an interesting formulation to study
thermal field theories on the lattice. There are 
many applications that can potentially benefit from 
them. In this section we sketch a few examples 
with the computation of thermodynamic potentials in mind.

\subsection{Temperature scan at fixed lattice spacing}
\FIGURE[t]{
\centerline{\includegraphics[width=10.0 cm,angle=0]{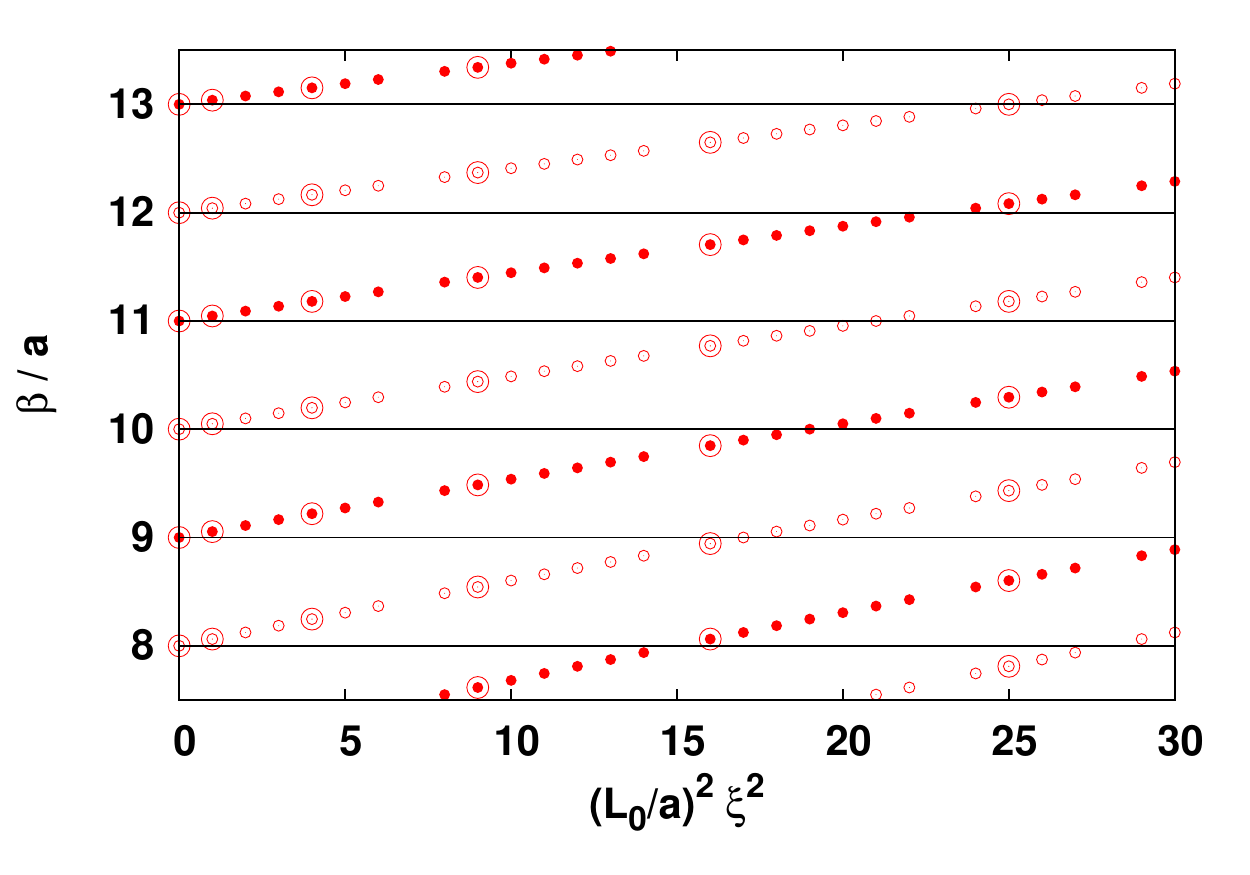}}
\vspace{-1cm}
\label{fig:varytemp}
\caption{Inverse temperature values that become accessible with the use
of shifted boundary conditions at a fixed lattice spacing $a$ and for 
different values of $L_0/a$. The inverse temperatures accessible with a 
shift in a single direction, $\vec \xi = (\xi_1,0,0)$, are marked by 
a double circle.}}
The possibility of varying the temperature by changing
either $L_0/a$ or $\vec\xi$ allows for a fine scan of the
temperature axis at fixed lattice spacing. This is 
illustrated in \fig\ref{fig:varytemp}, where it is also 
compared with the standard procedure of varying $L_0/a$ only.
This fact may turn out to be useful in all those cases where 
the temperature needs to be changed in small steps, e.g.
study of phase transitions etc.
\vspace{-0.25cm}

\subsection{Renormalization of the energy-momentum tensor}
\vspace{-0.25cm}
In the continuum, the charges associated with translational symmetries, i.e. 
the total energy and momentum fields, do not need any
ultraviolet renormalization thanks to the Ward identities 
that they satisfy, for a recent discussion see 
Ref.~\cite{Giusti:2011kt} and references therein. On the lattice, 
however, translational invariance is broken down to a discrete 
group and the standard charge discretizations acquire 
finite ultraviolet renormalizations. The energy-momentum field 
$T_{\mu\nu}$ is a symmetric rank-two tensor. Its traceless part is 
an irreducible representation of the SO(4) group. On the lattice, however, 
the diagonal and off-diagonal components of this multiplet  
belong to different irreducible representations of the hypercubic lattice 
symmetry group and therefore renormalize in a different way.
In SU($N$) Yang--Mills theory, they both renormalize multiplicatively.
The WIs can be enforced on the lattice to compute the 
overall renormalization constant $Z_T$ of the multiplet, and the relative 
normalization $z_{_T}$ between the off-diagonal and the diagonal 
components~\cite{Caracciolo:1989pt},
\be
T_{01}^{\rm R} = Z_{_T} T_{01},\qquad \qquad  
T_{00}^{\rm R}-T_{11}^{\rm R} = Z_{_T} \,z_{_T} \,(T_{00}-T_{11}), \qquad {\rm etc.}
\ee
where the fields with a superscript `R' are the renormalized ones.
There are many ways to implement this strategy in practice. 
A possible choice is to require a primitive matrix
\be\displaystyle
V_T = \left(\begin{array}{c@{~~}c@{~~}c@{~~}c}
L_0 &  0  & 0 & 0  \\
\frac{L_0}{2} & \frac{5}{2} L_0 & 0 & 0 \\
0  & 0 & L & 0 \\
0  & 0 & 0 & L 
\end{array}   \right)\; 
\ee
and compute $z_{_T}$ and $Z_T$ as (see also Ref.~\cite{Robaina:2013zmb})
\be
z_{_T} = \frac{3}{2} 
\frac{\langle T_{01} \rangle_{V_T}}
{\langle T_{00} \rangle_{V_T}  - \langle T_{11} \rangle_{V_T}}\; ,\qquad
Z_T = \frac{1}{L_0 \langle T_{0k} \rangle_{V_T}} 
\frac{\partial}{\partial \xi_k} \ln Z(V_T)\; . 
\ee
Alternatively $Z_T$ can be determined from ($x_0\neq y_0$, $x_2\neq y_2$)
\be\label{eq:bellissima5}
\frac{Z_T}{z_{_T}}=
\frac{\langle T_{00}\rangle_{V_T} -\langle T_{22}\rangle_{V_T}}
{L_0\,\langle {\overline T}_{02}(x_0)\, T_{02}(y)\rangle_{V_T, c} - 
L\,\langle {\widetilde T}_{02}(x_2)\, T_{02}(y)\rangle_{V_T, c}} 
\; .
\ee
Being fixed by WIs, the finite renormalization constants 
$Z_T$ and $z_{_T}$ depend on the bare coupling constant only.
Up to discretization effects, they are independent of the kinematics 
used to impose them, e.g. the volume, the temperature, the shift 
parameter, $x_0$ etc. Ultimately which WIs and/or kinematics 
yield the most accurate results must be investigated numerically.

\subsection{Calculation of the entropy and specific heat}
\vspace{-0.25cm}

Once the relevant renormalization constants are determined, 
the entropy density can be computed from the 
expectation value of $T_{0k}$ on a lattice with shifted boundary 
conditions,
\be\label{eq:rotThMuNu}
s = - \frac{Z_{T} L_0 (1+{\vec \xi}^2)^{3/2}}{\xi_k}\, 
\langle T_{0k} \rangle_{\vec\xi}\;, \qquad \xi_k \neq 0\; ,
\ee
by performing simulations at a single inverse temperature value 
$\beta=L_0\sqrt{1 + \vec\xi^2}$, and at a volume large 
enough for finite-size effects to be negligible. The latter are 
exponentially small in $(M L)$, where $M$ is the lightest 
screening mass of the theory \cite{Giusti:2012yj}. For the theory 
discretized with the Wilson action and for the `clover' form of the 
lattice field strength tensor, discretization effects turn out 
to be remarkably small \cite{Giusti:2012yj,Giusti:2013pepe}.
Once the entropy has been computed at various values of $\beta$,
the pressure can be computed by integrating $s$ in the temperature.
The ambiguity left due to the integration constant is consistent
with the fact that $p$ is defined up to an arbitrary 
additive renormalization constant.

The entropy density could also be computed  directly from 
\eq(\ref{eq:stwopt}) without the need for fixing the multiplicative renormalization 
constant. This would require, however, the computation of the two-point 
correlation functions in a large volume. The latter can also be used to access the 
specific heat of the system by using Eq~(\ref{eq:cvtwopt}).

We thank M. Pepe and D. Robaina for interesting discussions. This work was 
partially supported by the \emph{Center for Computational 
Sciences in Mainz}, by the DFG grant ME 3622/2-1 \emph{Static and dynamic
properties of QCD at finite temperature}, by the MIUR-PRIN contract 20093BMNPR,
and by the INFN SUMA project.

\end{document}